\documentstyle[twocolumn,aps,prb]{revtex}

\begin{document}
\draft

\twocolumn[\hsize\textwidth\columnwidth\hsize\csname@twocolumnfalse\endcsname%

\title{Logarithmic temperature dependence of conductivity at
half-integer filling factors:  Evidence for interaction between
composite fermions}

\author{L. P. Rokhinson, B. Su, and V. J. Goldman}

\address{Department of Physics, State University of New York,
Stony Brook, NY 1974-3800}

\date{July 14, 1995}

\maketitle

\begin{abstract}

We have studied the temperature dependence of diagonal conductivity in
high-mobility two-dimensional samples at filling factors $\nu=1/2$ and
3/2  at low temperatures.  We observe a logarithmic dependence on
temperature, from our lowest temperature of 13 mK up to 400 mK.  We
attribute the logarithmic correction to the effects of interaction
between composite fermions, analogous to the Altshuler-Aronov type
correction for electrons at zero magnetic field.

The paper is accepted for publication in Physical Review B, Rapid
Communications.

\end{abstract}


\vskip2pc]

Recently, an elegant theory of the fractional quantum Hall effect
(FQHE), including even-denominator filling factors, came with the
theory of composite fermions.[1,2]   In this theory, the system of
strongly interacting electrons can be mapped onto the system of
non-interacting new particles, composite fermions (CF), by attaching an
even number 2$m$ of vortices to each electron.  In the mean-field
approximation, the gauge field of vortices partially compensates the
external magnetic field $B$, and CF experience an effective magnetic
field $B^{cf} = B - 2mn\phi_0$, where $n$ is the concentration of
electrons (and CF), $\phi_0$ is the flux quantum and $m$ is an integer.
For $m=1$, at filling factor $\nu=1/2$, the external field is fully
canceled by the gauge field and $B^{cf}=0$.  It has been shown both
theoretically [2] and experimentally [3-5] that some properties of a
Fermi liquid are preserved for CF, in particular, a reasonably well
defined Fermi surface.

It is well known that at $B=0$, as temperature
$T$ is lowered, the diagonal conductivity $\sigma_{xx}$ of a disordered
two-dimensional electron system (2DES) nearly saturates, and only a
weak, logarithmic $T$-dependent correction to $\sigma_{xx}$ is
observed. Both weak localization and electron-electron interaction
effects in a disordered 2DES [6,7] give rise to logarithmic corrections
to conductivity at $B=0$. It has been suggested [2] that interactions
between CF could lead to a logarithmic corrections at $\nu=1/2$ also.
(Note that weak localization should be suppressed at $\nu=1/2$ because
gauge field fluctuations break time-reversal symmetry for impurity
scattering.)

In this paper we report the observation of a logarithmic correction to
conductivity of composite fermions at $\nu=1/2$ and 3/2.  This is, to the
best of our knowledge, the first observed effect that requires
interaction between CF.  We find that the coefficient of the
logarithmic term is significantly greater than that for electrons at
$B\approx 0$; the additional nonuniversal contribution may be
attributed [2] to the short-range interaction via gauge field
fluctuations.

We have studied several samples fabricated from high
mobility ($\mu\sim 2\times10^6\text{ cm}^2\text{/V s}$) GaAs/AlGaAs
heterojunction wafers.  The wafers have double Si $\delta$-doping, the
first layer is separated from the 2DES by a $d_s\approx 120$ nm thick
spacer.  2DES with densities between 0.4 and $1.2\times10^{11}
\text{cm}^{-2}$ were prepared by illuminating a sample with red light.
The temperature was measured with a calibrated Ruthenium Oxide chip
resistor.  Measurements were done in a top-loading into mixture
dilution refrigerator using standard lock-in technique at 2.5 Hz and
applied current 50 pA rms; no heating effects were observed at this
current.

We made samples in a Corbino geometry defined by circular In:Sn Ohmic
contacts with the inner radius $0.2\leq r_i\leq0.6$ mm and the outer
radius $r_o=1.5$ mm.  In the Corbino geometry, the local conductivity
is inversely proportional to the directly measured two-terminal
resistance $R_{2\text{T}}$: $\sigma_{xx}=\Box/R_{2\text{T}}$, where
$\Box=(\pi/2)\ln(r_o/r_i)$ is a geometric factor ("the number of
squares").  Representative magnetoconductivity data around $\nu=1/2$
and 3/2 are plotted in Fig.  1.  In this paper we concentrate on the
temperature dependence of the conductivity at half-integer filling
factors.  The data was collected either from the $B$-sweeps at several
temperatures, or by changing $T$ at a fixed magnetic field.  We checked
carefully that the sample and the thermometer were in close thermal
equilibrium during the measurement:  values obtained from both types of
measurements differ less than 0.5\% for the same ($B, T$).  As shown in
Fig. 2, in the experimental temperature range 13 mK $< T <$ 1.6 K,
$\sigma_{xx}$($\nu=1/2$) has a weak $T$-dependence with a minimum at
$\sim 500$ mK.  A remarkable result is that $\sigma_{xx}$ changes
logarithmically with $T$ at lower temperatures.

The variation of  $\sigma_{xx}$($\nu=1/2$) is less than 20\% over some
two orders of magnitude in $T$, this indicates a metallic conduction
regime.  At the same time, the conductivity is about 0.01 $e^2/h$, that
is, well below the Mott's minimum metallic conductivity for two
dimensions.  This apparent contradiction is resolved easily within the
theory of composite fermions.  Indeed, the theory [2] predicts a
metallic state for CF near $\nu=1/2$.  At $\nu=1/2$, CF experience zero
effective magnetic field, $B^{cf}=0$, and behave similar to electrons
at $B=0$.  This has been confirmed experimentally by observations of a
Fermi surface of CF in a weak effective magnetic field near
$\nu=1/2$.[3-5] The relationship between the transport coefficients of
the electron and CF systems were derived in Ref. 2, Appendix B;
specifically, the conductivity of composite fermions   can be
calculated from the measured electron conductivity $\sigma_{xx}$.
Since $\rho^{cf}_{xx}=\rho_{xx}$ and $\rho^{cf}_{xy}(\nu=1/2)=0$, at
$\nu=1/2$
\begin{equation}
\sigma^{cf}_{xx}=(1/\rho^{cf}_{xx})={{\sigma^2_{xx} +
\sigma^2_{xy}}\over{\sigma_{xx}}}.
\end{equation}

$\sigma^{cf}_{xx}(\nu=1/2)$ is plotted as a function of log$T$ for
two samples in Fig.  3.  Note that $\sigma^{cf}_{xx}$ is much greater
than $e^2/h$, as expected for a metallic regime.  Because the
$T$-dependent correction to $\sigma_{xx}$ is relatively small, the
functional dependence of the logarithmic correction is preserved after
the transformation from $\sigma_{xx}$ to $\sigma^{cf}_{xx}$ via Eq. 1,
although the sign is changed.  In all measured samples,
$\sigma^{cf}_{xx}$ shows a logarithmic temperature dependence for a
variation of $T$ by at least a factor of 30:
\begin{equation}
\label{scf-t}\sigma_{xx}^{cf}=\sigma_{0}^{cf} + \lambda {\frac{{e^2}}{{h}}}%
\ln{\frac{T}{{T_0}}},
\end{equation}
Here $T_0$ is a characteristic temperature (inverse of the elastic
lifetime [7]) and $\lambda$ is a positive dimensionless coefficient.
We find that experimental $\lambda$ is in the range
$0.4\leq\lambda\leq1.6$ and is sample-dependent.

Pursuing the analogy
with 2DES at zero field, the logarithmic correction is expected to be
of the Altshuler-Aronov type,[6] arising from the interaction effects
between CF in the presence of disorder.  At $\nu=1/2$ the
situation is more complex than at $B=0$ because CF interact
additionally with the fluctuations of the gauge field, and $\lambda$
has not been evaluated theoretically for this case.[2] Nevertheless, we
note that the Coulomb exchange contribution of $1/2\pi\approx0.32$ to
$\lambda$ is not sufficient to account for the high values observed.
For electrons at low $B$, for comparable GaAs 2DES samples, Paalanen et
al.  [8] obtained $\lambda\approx0.09$ and Choi et al. [9] obtained
$\lambda\approx0.17$ (note that CF at $\nu=1/2$ are
spin-polarized unlike the electrons at $B=0$).  CF experience
significant large-angle scattering on gauge-field fluctuations, in
contrast to the small-angle scattering dominant for electrons at zero
field.[2,10]  The short-range interaction between CF via gauge field
fluctuations may lead to an additional non-universal contribution to
$\lambda$.

CF can also be formed in the spin-split Landau level.  At $\nu=3/2$ the
spin-up ($\uparrow$) level is fully occupied by (2/3)$n$ electrons and
(1/3)$n$ electrons occupy the spin-down ($\downarrow$) level to one-half
filling.  Assuming that both spin-split levels are decoupled, the total
conductivity tensor can be written as [11]:
\begin{equation}
\label{s-total}\sigma=\sigma^\uparrow + \sigma^\downarrow,
\end{equation}
where
$$
\sigma^{\uparrow} ={\frac{{e^2}}{{h}}}\left[
\begin{array}{cc}
0 & 1 \\
-1 & 0
\end{array}
\right]
$$
and $\sigma^{\downarrow}$ is similar to $\sigma_{xx}$ for $\nu=1/2$.
Thus, $\sigma^{cf}_{xx}$ for $\nu=3/2$ can be calculated by
substituting $\sigma^{\downarrow}_{xx}$ instead of $\sigma_{xx}$ into
Eq. (1).  The resulting $\sigma^{cf}_{xx}(\nu=3/2)$ is approximately
(1/3) $\sigma^{cf}_{xx}(\nu=1/2)$ because the density of CF is reduced
3 times at $\nu=3/2$, compared to $\nu=1/2$.  Experimentally, a factor
of 3 difference between $\sigma_{xx}$ (or $\sigma^{cf}_{xx}(\nu=1/2)$ )
at $\nu=1/2$ and 3/2 is observed in the full experimental temperature
range.  In Fig.  3(a), the left and right axes (for $\nu=3/2$ and 1/2,
respectively) differ by a factor of 3 to emphasize this fact.

At $\nu=3/2$,  $\sigma^{cf}_{xx}$ also has a logarithmic temperature
dependence at low $T$ [Fig.  3(a)], although the experimental range of
temperatures is limited to 13 - 180 mK.  The similarity of the
$T$-dependencies of $\sigma^{cf}_{xx}$ at $\nu=1/2$ and 3/2 suggests
that the underlying physics is the same and originates from the
interactions between CF in the spin-split Landau level.  The
coefficient $\lambda$, obtained from a fit of the data with Eq. 2, is
about four times smaller than at $\nu=1/2$.  It can be expected that
the screening length at $\nu=3/2$ is smaller compared to $\nu=1/2$ (2/3
of electrons are in the fully filled spin-split level and therefore
cannot screen disorder potential).  However, since no explicit
theoretical treatment of the logarithmic correction to the conductivity
of CF exists at present, it is not clear what can affect $\lambda$ in
this regime.

One of the observable consequences of the electron interaction effects
at $B\approx0$ is a temperature-dependent parabolic negative (for
$\lambda>0$) magnetoresistance.[12]  At $\nu\approx1/2$, the CF
resistivity follows the same $B$-dependence as $\sigma_{xx}$ (see Fig.
1).  Thus, in contrast to the $B\approx0$ case, in the vicinity of
$\nu=1/2$ magnetoresistance is positive, despite the fact that the
measured $\lambda>0$.  Presumably, the different sign of the
magnetoresistance at $B\approx0$ and $\nu\approx1/2$ originates from
the difference in  leading scattering mechanisms in two regimes.
Electrons at $B\approx0$ predominantly scatter on fluctuations of
electron density $\delta n$ that are a reflection of a random
distribution of remote donors.  These fluctuations should be of the
order of one per cent of the total electron concentration $n$.  At
$\nu\approx1/2$, the fluctuations of the electron density produce
fluctuations of the effective magnetic field $B^{cf}$ for CF: $\delta
B^{cf}/B_{1/2}=\delta n/n$. For a typical magnetic field
$B_{1/2}\sim10$ T and $\delta n/n\sim1\%$, fluctuations of
$B^{cf}$ can be as big as 0.1 T.  Thus, the dominant scattering
mechanism for CF is on fluctuations of the effective magnetic field
that may account for high experimental resistance,[2] large-angle
scattering,[13] and, perhaps, the positive magnetoresistance at
$\nu\approx1/2$.  To support this conjecture we would like to note
that a spatially nonuniform static magnetic field leads to a positive
magnetoresistance at $B\approx0$.[14]

Previous experiments, focused on the study of the FQHE states or the
existence of the Fermi surface for CF around $\nu=1/2$, are
sufficiently well described by theories of non-interacting CF.[1] We
have found that a simple picture of non-interacting CF is not fully
adequate, and that the CF interaction effects should be taken into
account in a more comprehensive description of the FQHE regime.  This
finding further deepens the similarity between 2DES at $B\approx0$
and at $\nu\approx1/2$.

We are grateful to M. Shayegan for MBE material and to S.  Sondhi, J.
K.  Jain, P.  A.  Lee and D. B.  Chklovskii for interesting
discussions.  This work was supported in part by NSF under Grant No.
DMR-9318749.

\bibliographystyle{prsty}

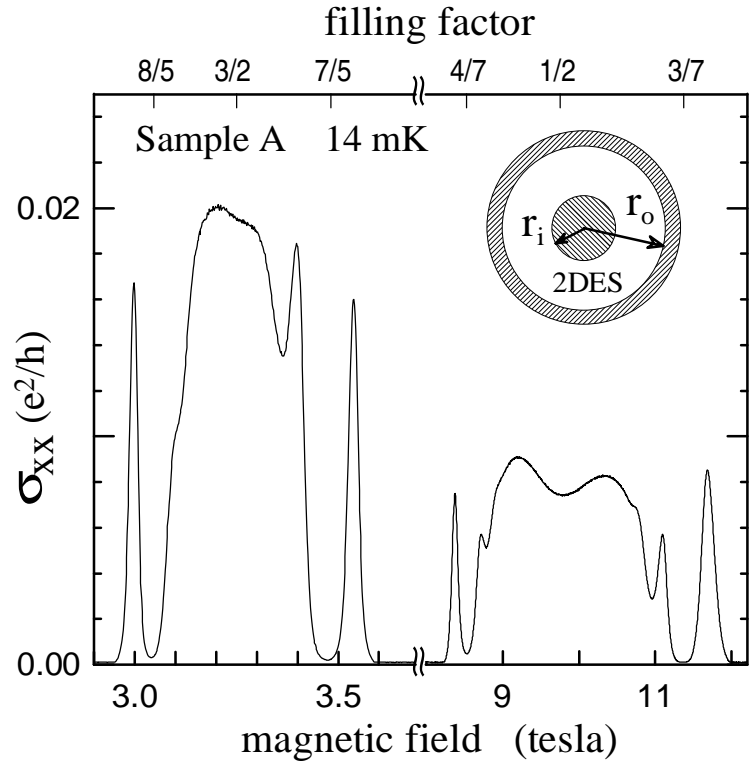
Fig. 1. The electron diagonal magnetoconductivity of Sample A near $\nu=1/2$
(right) and 3/2 (left) at $T$=14 mK; note the break in the horizontal axis.
The inset shows the sample layout in the Corbino geometry.

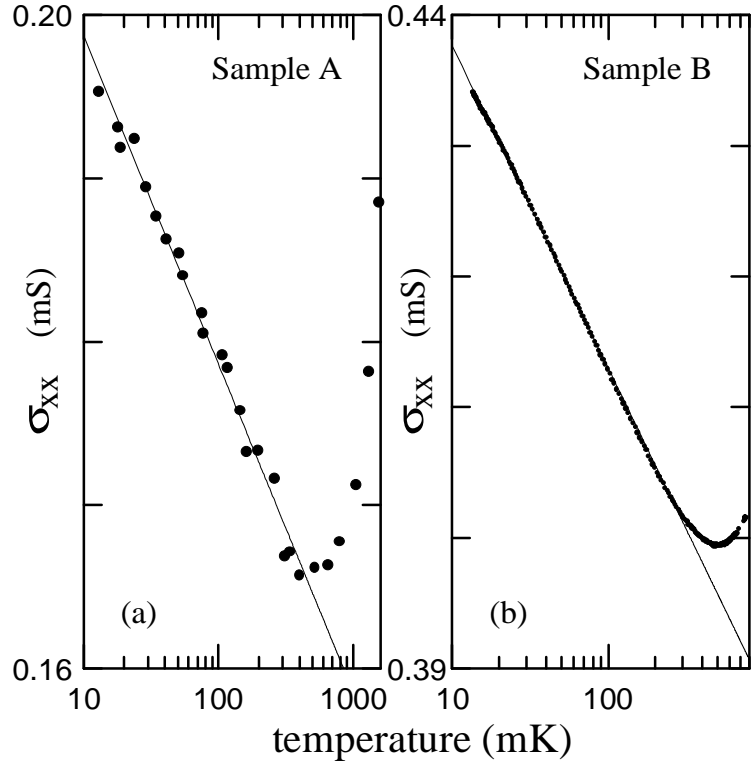
Fig. 2.  The temperature dependence of the electron diagonal conductivity
at $\nu=1/2$ in samples A and B.  The straight lines are the logarithmic
fits for the temperature range from 15 to 300 mK.

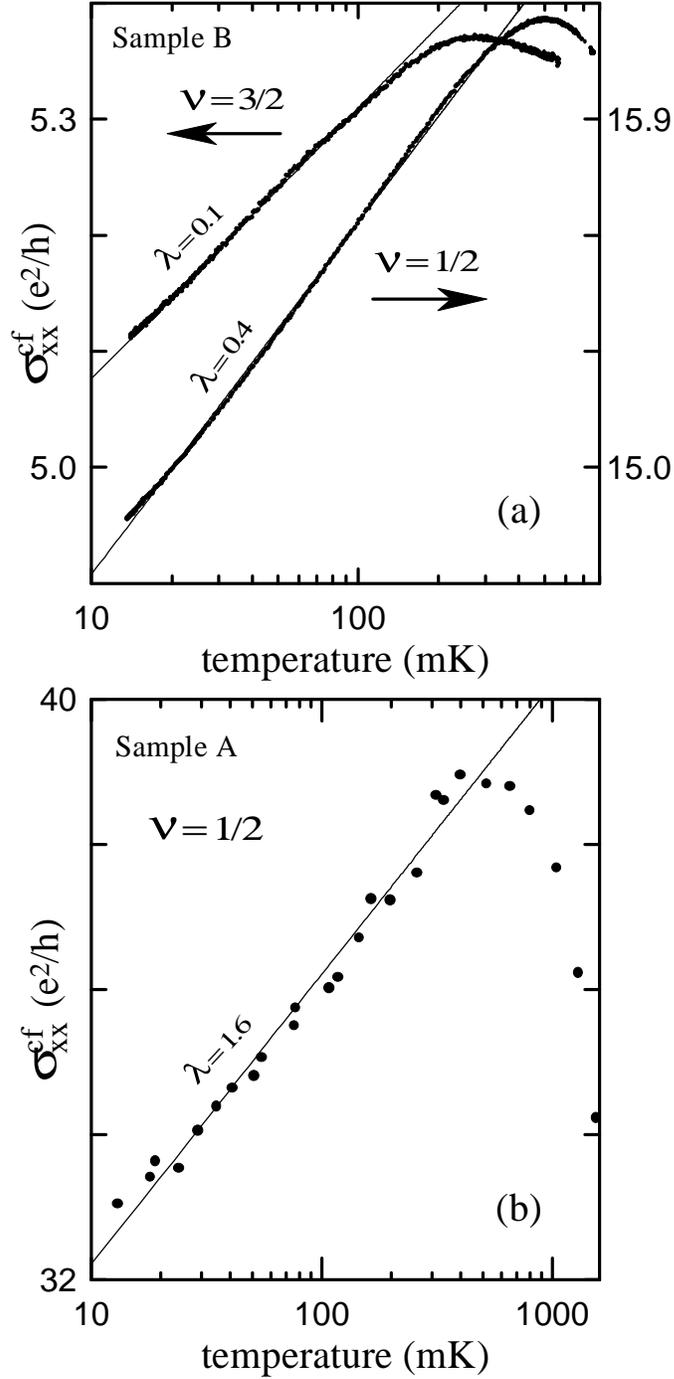
Fig. 3.  The temperature dependence of the composite fermion (CF)
conductivity for the same samples as in Fig. 2.  The vertical axes in
(a) differ by a factor of 3 for the $\nu=1/2$ and 3/2 data for sample B.
The slope of the fitted straight lines is $\lambda$, defined in Eq. (2).

\end{document}